\documentclass{PoS}

\newcommand{\rmn}[1]{{\mathrm{#1}}}
\newcommand{\la}{\,\rlap{\raise 0.5ex\hbox{$<$}}{\lower 1.0ex\hbox{$\sim$}}\,}

\newcommand\degr{\hbox{$^\circ$}}

\title{Results of INTEGRAL TOO observation of the transient X-ray burster XTE J1810-189}

\ShortTitle{TOO observation of XTE J1810-189}

\author{\speaker{I.V. Chelovekov}\\
        Space Research Institute, Russian Academy of Sciences, Moscow, Russia\\
        E-mail: \email{chelovekov@iki.rssi.ru}}

\author{S.A. Grebenev\\
        Space Research Institute, Russian Academy of Sciences, Moscow, Russia\\
        E-mail: \email{grebenev@iki.rssi.ru}}

\abstract{We report results of the INTEGRAL Target of Opportunity
observations of the transient X-ray burster XTE J1810-189.  The
observations were performed on April 3--6, 2008, soon after the
discovery of the source and near the peak of its outburst. That
time the source had a flux of about 50 mCrab and exhibited a
hard Comptonized X-ray spectrum extending well above 100
keV. Being approximated by a power law with an exponetial
cut-off in the broad 3--100 keV energy band it gave the average
photon index $\Gamma\simeq 1.6$ and $kT_{cutoff}\simeq 67$ keV. We found
only slight indications for changes in the index during the
observation ($\Gamma$ first steady decreased from $\sim 2.0$ to
$\sim1.3$ and then increased back to $\sim 2.0$). However the
$N_{\rmn H}$ value measured by absorption in the low energy part
of the spectrum changed drastically and very irregularly (from
$\sim 4\times 10^{22}$ till $\sim 100\times 10^{22}$ cm$^{-2}$).
There were 10 type I X-ray bursts detected from the source
during these TOO observations. Assuming that the Eddington
luminosity was reached during the burst with the highest peak
flux we get an upper estimate for a distance to the source
$D=6.4\pm0.6$ kpc. From the X-ray burst parameters we conclude
that this LMXB harboures an evolved star.}

\FullConference{The Extreme sky: Sampling the Universe above 10 keV - extremesky2009,\\
		October 13-17, 2009\\
		Otranto (Lecce) Italy}

\begin{document}

\section{Introduction}
The X-ray transient XTE~J1810-189 was discovered in the spring 
of 2008 during RXTE/PCA monitoring scans of the Galactic ridge
region \cite{markwardt08}. A pointed observation started on
March 10, at 21:05 UTC revealed a variable source (30\%
r.m.s. fluctuations). The emission spectrum was
consistent with an absorbed power law ($N_{\rmn
H}\simeq1\times10^{22}~\mbox{cm}^{-2}$, photon index
$\Gamma\simeq1.9$). The PCA flux history suggested its gradual rise
since March 5. The source was observed on March 12-15
with INTEGRAL that measured a slightly steeper spectrum
$\Gamma=2.26\pm0.12$ in the
hard $>20$ keV IBIS/ISGRI energy band 
\cite{neronov08}. The observation of XTE~J1810-189 on March
17, 2008 with the Swift/XRT telescope revealed the similar
spectrum but the higher absorption $N_{\rmn
H}=(4.2\pm0.7)\times10^{22}~\mbox{cm}^{-2}$ indicating that
there might be an internal source of absorption in the system
\cite{krimm08}. In a pointed observation of the
source on March 26, at 12:47 UTC the RXTE/PCA detected a type I
X-ray burst identifying a compact object in the system as
a neutron star. Assuming the Eddington peak luminosity, the
upper limit  for a distance to the source was
obtained $D\la 11.5$~kpc \cite{markwardt08b}. 

In this paper we report the results of TOO (Target of
Opportunity) observations of XTE J1810-189 performed with
INTEGRAL on April 3-6, 2008.

\section{Observations and data analysis}

Based on the detection of the outburst from XTE J1810-189 in
2008 March and its rising X-ray brightness it was decided to perform
a TOO observation of the source with INTEGRAL. All together there were 
57 individual pointing observations (43 in revolution 668 and 14 in 
revolution 669), each lasting 1--1.5 hours. The total exposure during 
this TOO observation was over 203 ks.

The data set used in this research was obtained by the ISGRI (an upper
detector of the IBIS telescope) and by the Module 1 of the JEM-X telescope
aboard INTEGRAL. For detailed description of the IBIS/ISGRI and JEM-X 
telescopes see \cite{ubertini03,lebran03} and \cite{lund03} respectively.

Data reduction was performed using the OSA 7.0 software, distributed by
the Integral Science Data Center (ISDC). In order to search for X-ray
bursts we analyzed the JEM-X (3--20 keV) and IBIS/ISGRI (15--25 keV) detector 
light curves (count rate histories), using all the detector events, 
irrespective to their incidence directions. See \cite{cgs06}
for more detailed description of the technique.

The spectral fitting was performed with Xspec package, version 11.3.1.

\section{Results}

\begin{figure}
\centerline{\includegraphics[width=0.45\linewidth]{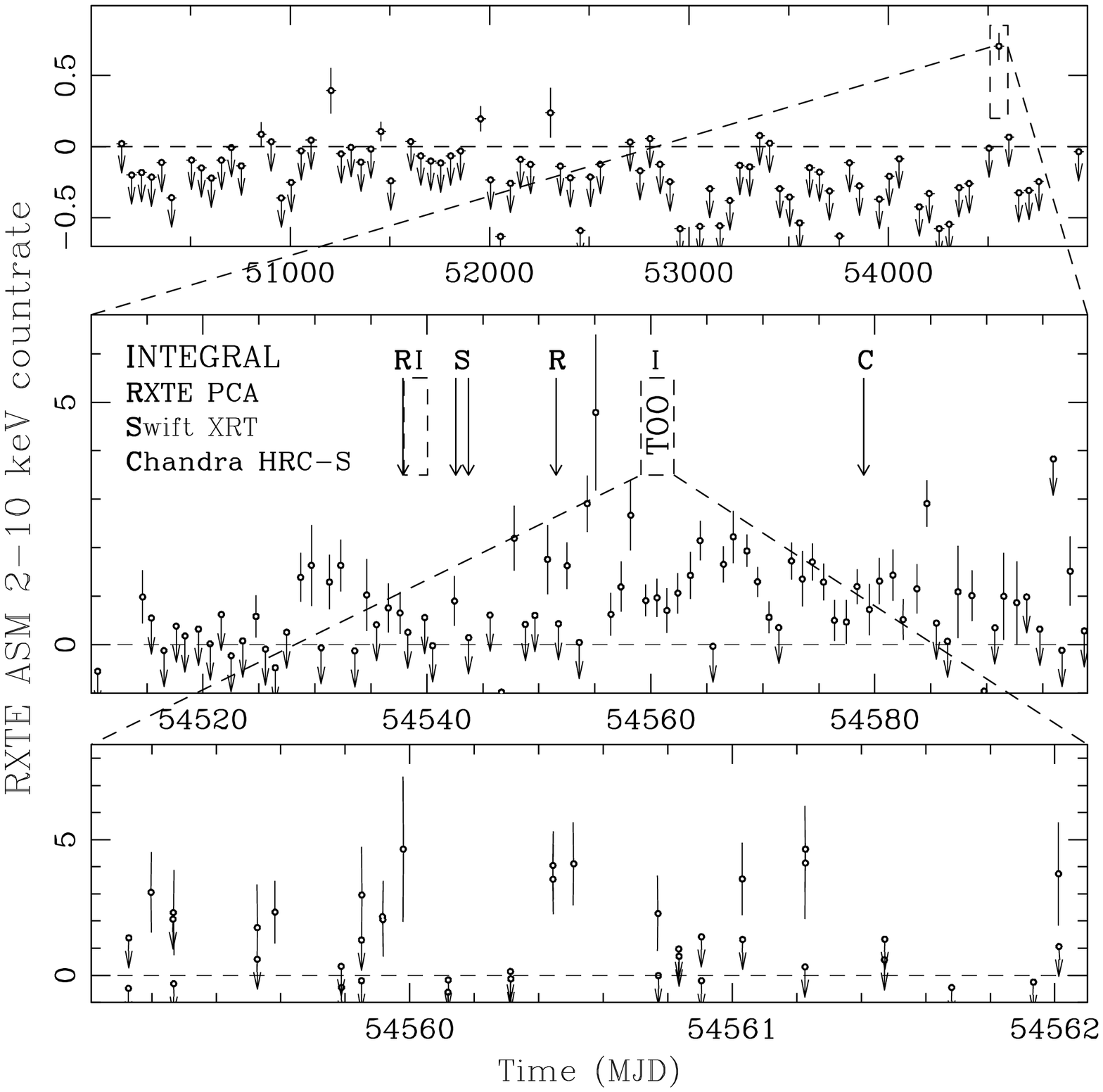}\includegraphics[width=0.45\linewidth]{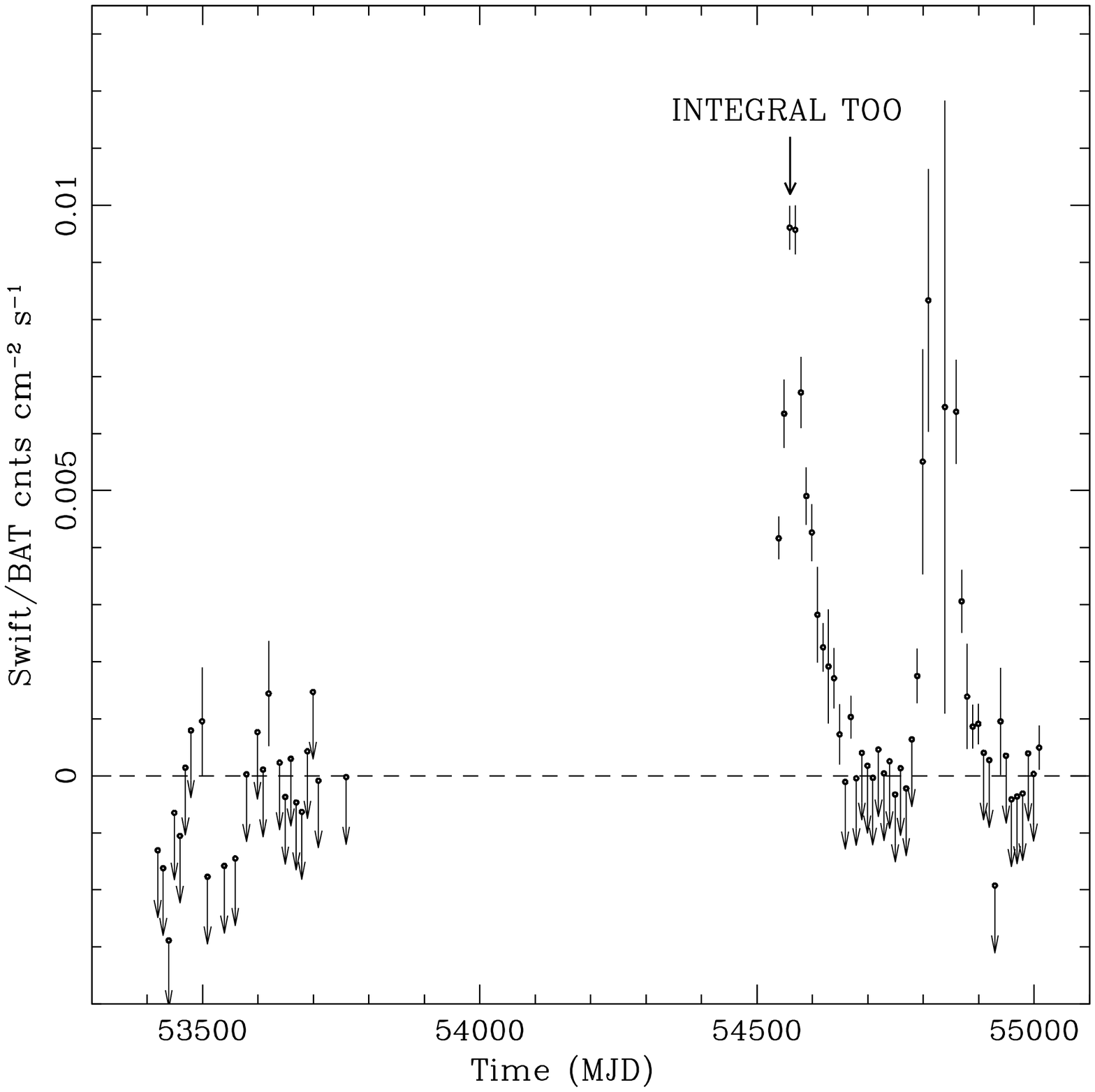}}
\caption{Left: All three panels show the RXTE/ASM count rate
from XTE J1810-189 in the 2--10 keV energy band.
Right: Swift/BAT count rate history of \mbox{XTE 1810-189}
in the 15--50 keV energy band.}

\label{fig:asm_swift_lc}
\end{figure}
%-------------------------------------------------------------
Fig. \ref{fig:asm_swift_lc} (left) shows an extended ASM light curve of
XTE~J1810-189 observations (2--10 keV, top panel) note that
70 counts/s = 1 Crab $\simeq2.2\times10^{-8}~erg~s^{-1} cm^{-2}$. 
Each point of the top curve represents a 50 days averaged ASM 
count rate measured from XTE\,J1810-189. Here we omitted all the
bins, containing less then 15 observing days ($\le$30\% fill up). The
middle panel shows the same light curve with better resolution 
(each point corresponds to a 1 day averaged ASM count rate) and 
includes some history of the XTE J1810-189 discovery and observations. 
Note that the light curve may be contaminated by X-ray bursts represented 
by single point excesses (e.g. 54585), while count rate excesses consisting of 
several points (e.g. 54560-54570 MJD) are most likely due to the 
XTE J1810-189 flaring activity. The bottom panel is a yet more detailed 
ASM light curve, corresponding exactly to the interval of the INTEGRAL TOO 
observations. Each point on this panel represents already the
one-dwell ($\sim90$ s) averaged ASM count rate.

Fig.\,\ref{fig:asm_swift_lc} (right) shows the Swift/BAT count rate
history of XTE J1810-189 in the 15--50 keV energy band. Note, that 
in this energy band 1 $count~s^{-1} cm^{-2}$ = 4.6 Crab 
$\simeq1.4\times10^{-8}~erg~s^{-1} cm^{-2}$. An arrow 
marks the epoch of the INTEGRAL TOO observation. We can see that this 
INTEGRAL TOO observation took place during the peak of the 
source's hard X-ray outburst (note, that the ASM light curve shows 
that it was carried out $\sim$ 10 days before the peak in the 
standard X-ray band). The SWIFT/BAT energy band is not affected by interstellar 
or system inner absorption, so the outburst is clearly seen. From
this curve we can see, that the source recently underwent another 
outburst which started in December 2008 and continued
on till January 2009 ($\sim54800$--54880 MJD). The gap in the
curve in the middle of this outburst and a comparatively poor 
statistics are due to the lack of the
Swift/BAT data. The lack of observational data does not
allow one to empirically determine whether such outbursts are 
regular or the observed one is a standalone. This outburst may as well be 
an evidence for the beginning of an activity period, like the 
one observed from KS1731-294 in 1989-2001 (see \cite{cgs06}).

\subsection{Spectral analysis}

\begin{figure}
\centerline{\includegraphics[width=0.45\linewidth]{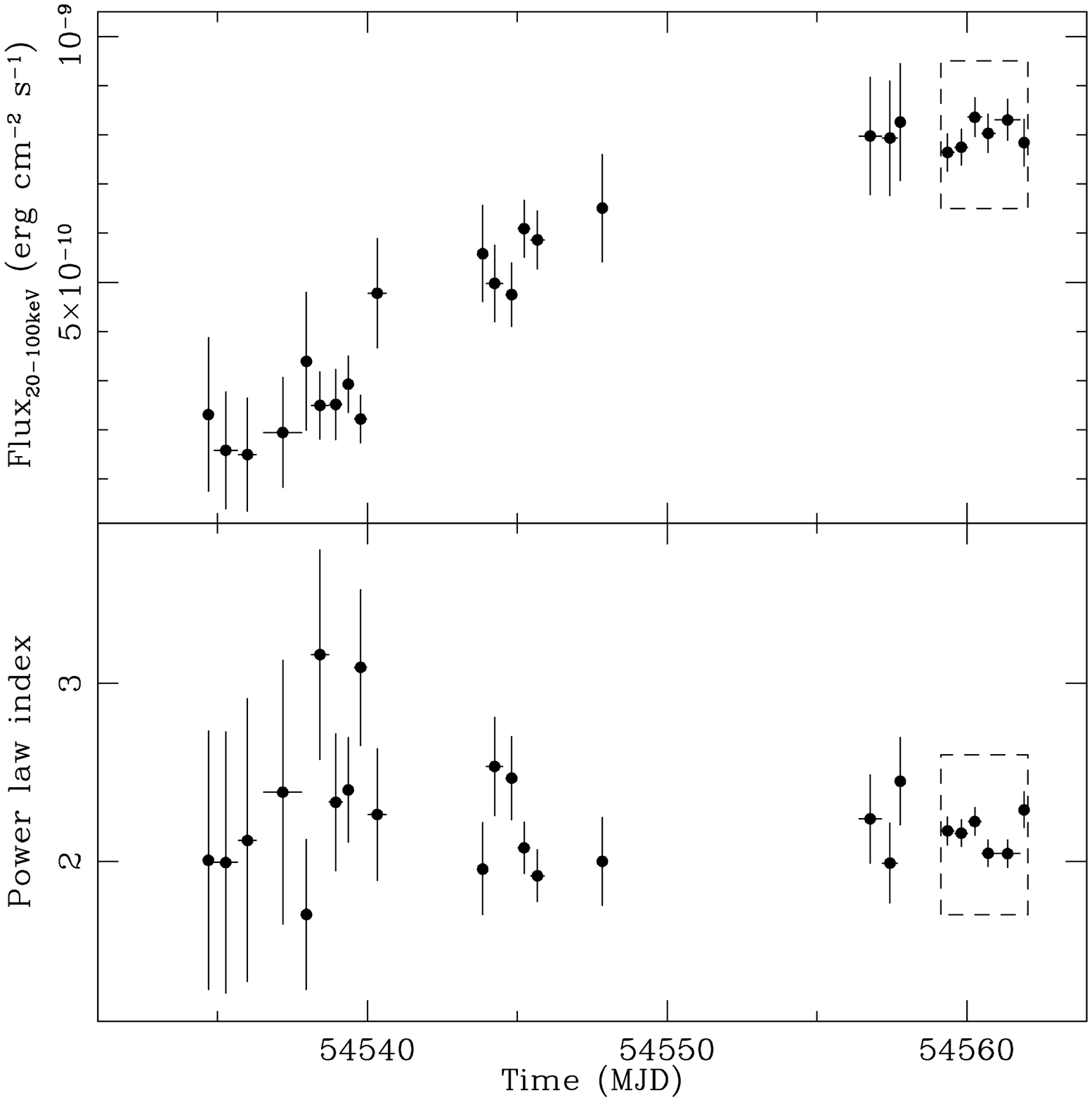}\includegraphics[width=0.45\linewidth]{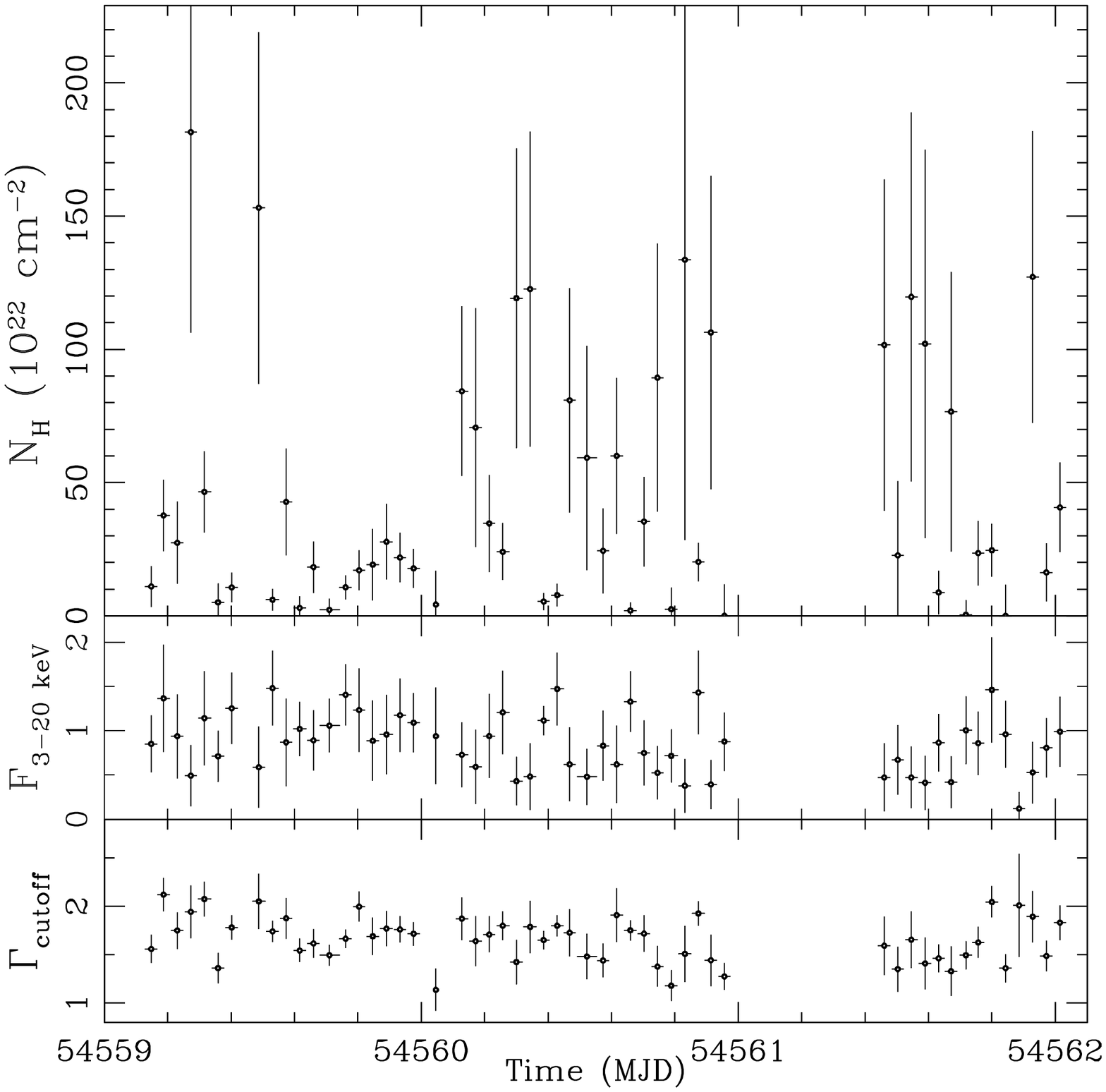}}
\caption{Left: Results of fitting the IBIS/ISGRI 20--100 keV
spectra with a power law model. 
Right: Results of fitting the broad band (3-100 keV) JEM-X+IBIS/ISGRI 
spectra with a cutoff power law model with interstellar absorption.}
\label{fig:spe_pars}
\end{figure}

In addition to the TOO observations, there were 229 science windows (SCWs)
with XTE\,J1810-189 inside the IBIS FOV, but none with the source
inside the JEM-X one. This let us examine the behavior of
the source in the hard 20--100 keV energy band. Fig.\,\ref{fig:spe_pars} 
(left) shows the results of fitting IBIS/ISGRI spectra of the source 
by a power law model. Each point in the figure corresponds to 10
individual INTEGRAL observations (SCWs). The upper panel shows
20--100 keV flux from the source. The lower panel shows a power law
index. Dashed boxes on each of the panels designate the INTEGRAL
TOO observations. Although there are some variations in the spectral
hardness during raising of the outburst, it shows no trend
correlation with the hard X-ray flux. The spectrum
remains Crab-like with an index $\sim2.0-2.5$ most of the time.

During the 5 INTEGRAL revolutions ($\sim$ 15 days) after the TOO
observation the source was
also often within the IBIS FOV (and again never in the JEM-X one)
but it was mostly over $10\degr$ off axis and less significant than
6--7 standard deviations, so we were not able to trace further
evolution and decay of the outburst.
%-------------------------------------------------------------

Fig.\,\ref{fig:spe_pars} (right) shows the results of fitting the broad
band (3--100 keV) JEM-X+IBIS/ISGRI spectra to a cutoff power law
model with interstellar absorption. Each point corresponds to one 
science window during the TOO observation. The upper panel shows a 
level of absorption in units of $10^{22}~cm^{-2}$. The middle panel shows the 
3-20 keV model flux from the source in units of $10^{-9}~erg~cm^{-2}~s^{-1}$.
The lower panel shows power law index. The cutoff energy was fixed at 
its average value $E_{cutoff}=66.7$ keV, obtained during fitting 
individual SCW spectra. The gap in the data is due to the INTEGRAL 
passing through Earth's radiation belts. Note, that the power law
index first gradually decreased from $\sim$2 to $\sim$1.3 and then 
came back to $\sim$2.0. The flux from the source in 20-100 keV did 
not vary with the 3-20 keV flux. This indicates that mass accretion 
rate or structure of inner accretion disk and area around boundary 
layer, where harder photons not affected by absorption are originated 
from, did not change. The cause of quasi-periodic variations of soft 
X-ray radiation (see Fig. \ref{fig:spe_pars} right) could be absorption 
in irregular structures at the outer edge of the accretion disk.

%-------------------------------------------------------------
\begin{figure}
\centerline{\includegraphics[width=0.55\linewidth]{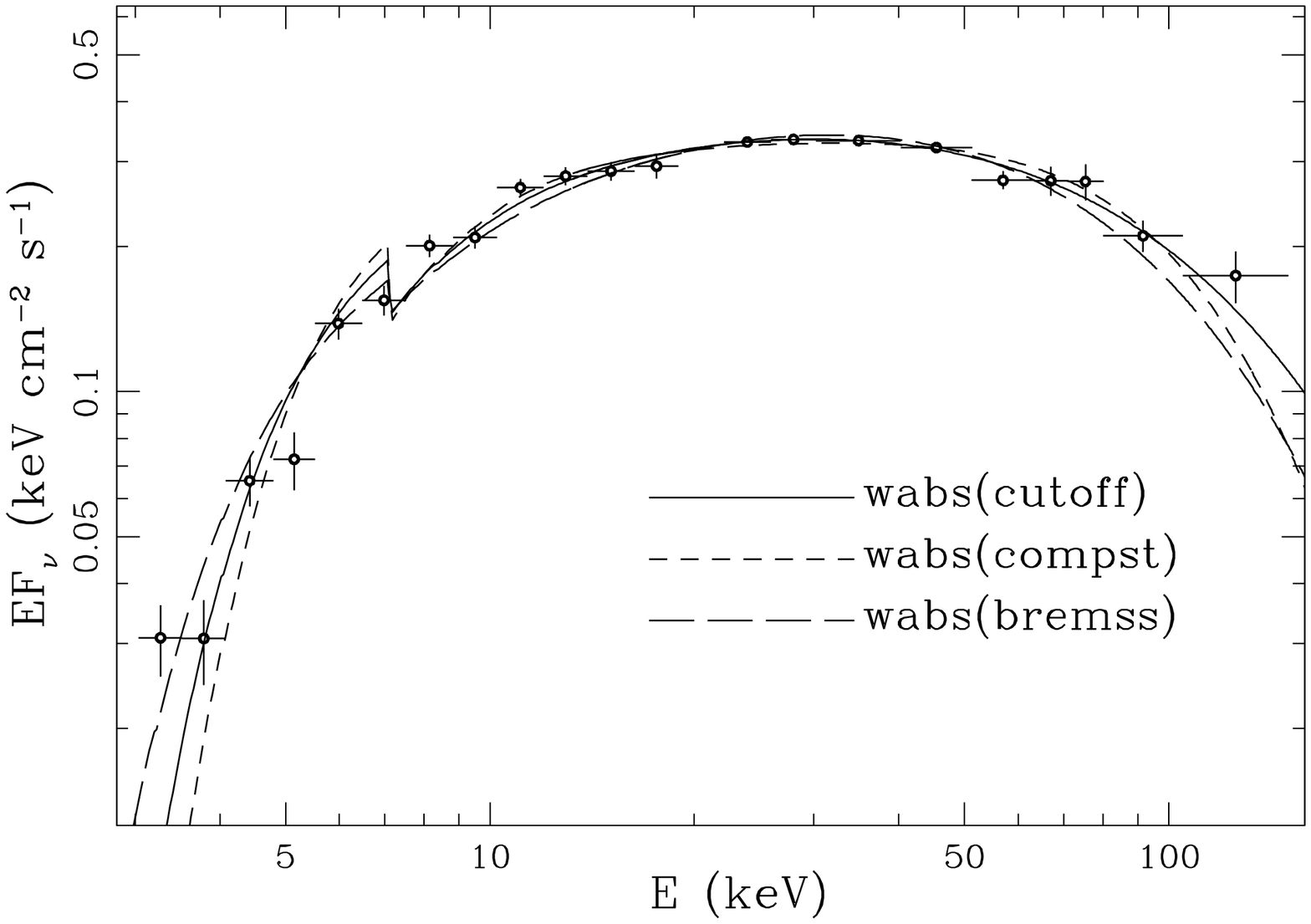}\includegraphics[width=0.40\linewidth]{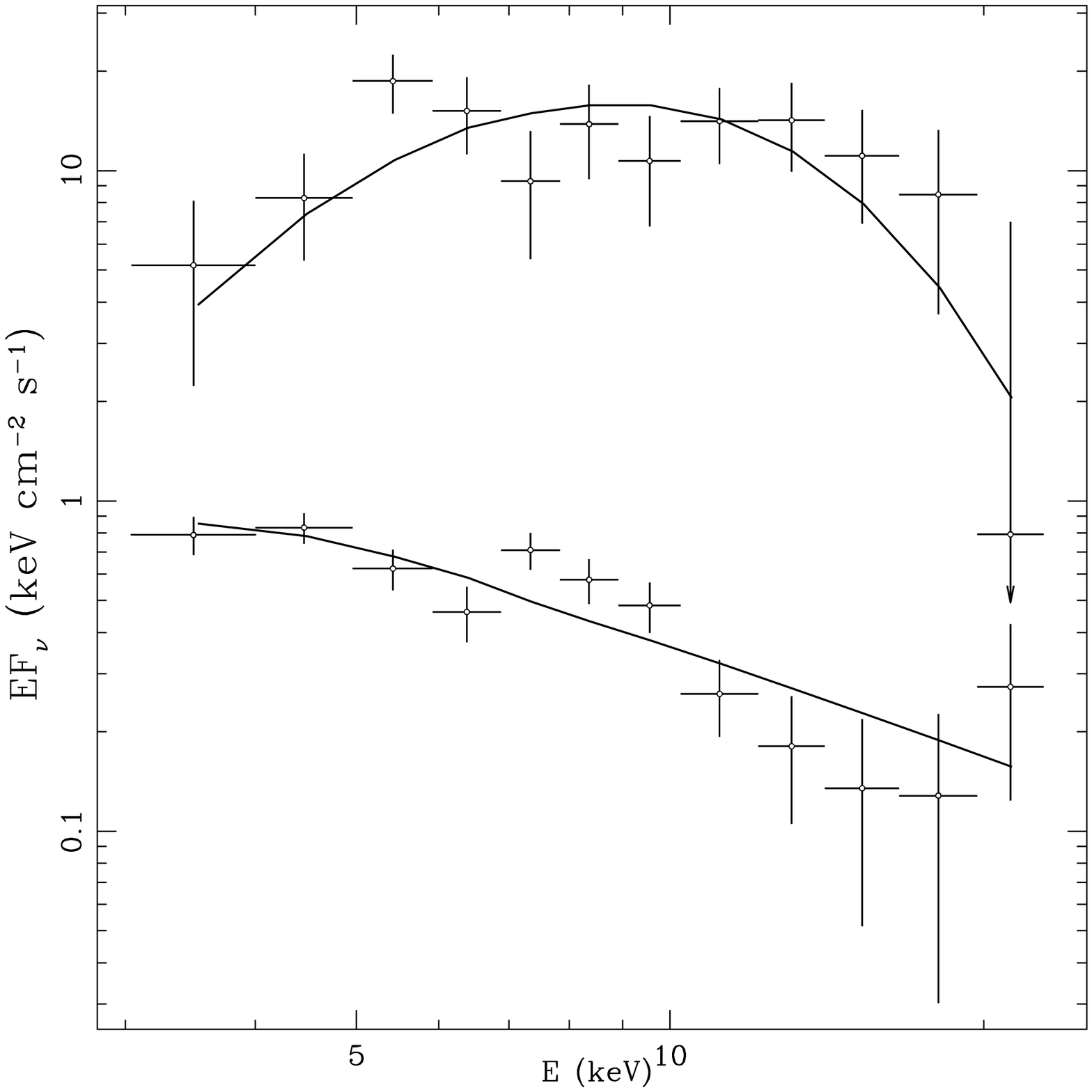}}
\caption{\rm Left: Average INTEGRAL TOO broad band JEM-X+IBIS/ISGRI spectrum of 
the XTE J1810-189 persistent emission.
Right: Spectrum of the X-ray burst detected from XTE J1810-189 at 10:38:30 
on Apr 4 (UTC) (top) and spectrum of the XTE J1810-189 persistent emission 
during the whole 1 hour INTEGRAL pointing observation, containing the burst 
(bottom).}
\label{fig:specs}
\end{figure}

\begin{table*}

\caption{Best-fit parameters for the broad band 3-100 keV JEM-X+IBIS spectrum of \mbox{XTE J1810-189} integrated over the whole set of TOO observations.}

\vspace{-0.5cm}

\begin{center} 
\begin{tabular}{ccccc}

\multicolumn{3}{c}{\it }\\
\hline
\hline

{Model}&
{$N_{H}$}&
{$\Gamma_{cutoff}/\tau_{compst}$}&
{$E_{cutoff}/kT_{bremss}/kT_{compst},$}&
{$\chi^{2}/N(N)$}\\
{}&
{$10^{-22}~cm^{-2}$}&
{}&
{keV}&
{}\\

\hline
\hline
$ wabs(cutoff) $ & $ 24.9\pm2.8 $ & $ 1.58\pm0.07 $ & $ 66.7\pm0.1 $ & $ 2.43(17) $ \\
$ wabs(bremss) $ & $ 15.0\pm1.4 $ & $ - $           & $ 49.0\pm1.4 $ & $ 2.97(18) $ \\
$ wabs(compst) $ & $ 36.4\pm2.5 $ & $ 3.7\pm0.2 $   & $ 23.0\pm1.4 $ & $ 3.32(17) $ \\

\hline
\end{tabular}
\end{center}

\vspace{-0.7cm}

\label{tab:pspec_pars}
\end{table*}

Fig.~\ref{fig:specs} (left) shows the average INTEGRAL TOO broad 
band (3--100 keV) combined JEM-X and IBIS/ISGRI spectrum of the source. 
JEM-X spectrum was renormalized during fitting by factor 1.25 (this factor 
was obtained from similar fitting of the Crab spectra).
Tab.~\ref{tab:pspec_pars} shows the results of fitting this spectrum
by several well known spectral models. Note that values of the reduced 
$\chi^{2}$ are quite high mostly due to JEM-X spectrum which is an average of
many non uniform individual SCW spectra. Fig.~\ref{fig:specs} (right)
shows the spectrum of the X-ray burst detected from XTE J1810-189 at 10:38:30 
on Apr 4 (UTC) and its Black Body fit ($kT_{BB}=2.2\pm0.2$ keV) (top) and 
spectrum of the XTE J1810-189 persistent emission during the whole 1 hour 
INTEGRAL pointing observation, containing the burst and its approximation by
a power law with $\Gamma=3.15\pm0.13$ (bottom).

The analysis of combined JEM-X+IBIS/ISGRI spectra shows that the source
had a rather hard spectrum during the whole TOO observation (at the peak of the 
outburst) which is expected at estimated 3-100 keV luminosity at the level 
$\sim$5-10\% of Eddington. The only variation of the spectrum is due to varying 
absorption, presumably due to the presence of complex structure at the outer 
edge of the disk. Irregular as these variations seem to be, yet at least during the
day 54560 (MJD) (see Fig.\ref{fig:spe_pars} right) the 3-20 keV flux varies with a 
distinct 5-6 hours period. Note, that generally flux rises when $N_{H}$ drops. At 
the same time harder X-ray radiation remains constant, so we can not attribute the
periodicity to dips caused by orbital eclipse.

%=============================================================
\subsection{X-ray bursts}

When the accretion rate in the transient binary star rises
during the outburst, one would naturally expect the type I X-ray
burst rate to also rise, because generally it should take a
shorter time to accumulate enough matter on the surface of the
neutron star for a thermonuclear explosion to ignite. Considering 
the fact, that the TOO observation was performed during
the peak of the flare, when the accretion rate is at
maximum, we made an effort to search for type I X-ray bursts in
JEM-X, as well as in IBIS data (many type I X-ray bursts are
also well seen in harder X-rays, see \cite{cgs07}). 
We first analyzed detector light curves of 57 TOO individual observations. 
JEM-X data revealed 10 significant type I X-ray bursts from 
\mbox{XTE J1810-189}. All the bursts had 3-5 s rise and 
10-15 s decay times. None of the observed bursts were strong enough 
to present an evidence of photospheric expansion witnessing of their 
Eddington luminosity. So we were only able to estimate an upper limit for the 
distance to the source using the $E_{Edd}=2.1\times 10^{38}~erg~s^{-1}$ 
(we assume the neutron star mass to be $1.4~M_{\odot}$) and the 
highest peak flux achieved $F_{peak}=4.3\times10^{-8}~erg~cm^{-2}~s^{-1}$ 
(second burst on Apr 4): D=$6.4\pm0.6$ kpc. 
This improves a previous estimation: D$\simeq11.5$ kpc (see \cite{markwardt08b}).

We note that the stability of recurrence time from burst to burst is remarkable.
The recurrence time is always within the limits of 5.3-6.1 hours. This
shows that accreting conditions in the system were quite stable during TOO
observations, which is expected at the top plateau of the outburst. 
Both comparatively high obtained values of $\alpha\sim80-250$ (ratio of
the gravitational energy released between the consequent bursts and the 
thermonuclear energy released during the burst - it is well over 40 which is 
expected for solar composition) and 
$\dot{m}\sim1.3-1.5\times 10^{3}~g~cm^{-2}~s^{-1}$ 
(accretion rate per surface area) are typical for pure He 
bursts (see \cite{bildsten2000}). But pure helium bursts should have recurrence 
times $\sim 10^{3}/Z_{CNO}$ s which is about one day for solar composition - 
time needed for stable burning of hydrogen. Here $Z_{CNO}$ is the mass 
fraction of CNO in accreted matter. The observed burst recurrence
time in the system is 5-6 hours. That implies a hydrogen poor matter with 
$Z_{CNO}\sim0.05$ and the donor to be an old evolved star. 

Fig.\ref{fig:spe_pars} (left) shows the rise of the outburst from XTE J1810-189. 
If we assume that the first several points represent the level of its persistent 
flux it is possible to predict the expected burst recurrence interval during the 
quiescent period, assuming the burst trigger conditions do not change: 
$\tau\sim30$ hours.

\end{document}